\documentclass[12pt]{iopart}
\usepackage{graphicx}
\begin{document}
\def \lt{\!\!<\!}

\title[Legacy data]{Legacy data and cosmological constraints from the 
angular-size/redshift relation for ultra-compact radio sources}

\author{J C Jackson and A L Jannetta\dag
\footnote[2]{To whom correspondence should be addressed
(john.jackson@unn.ac.uk)}}

\address{\dag\ Division of Mathematics and Statistics, School of Informatics,
Northumbria University, Ellison Place, Newcastle NE1 8ST, UK}

\begin{abstract}

We have re-examined an ancient VLBI survey of ultra-compact radio sources at 2.29 GHz,
which gave fringe amplitudes for 917 such objects with total flux density $^>_\sim~0.5$ Jy.
A number of cosmological investigations based upon this survey have been published in
recent years.  We have updated the sample with respect to both redshift and radio information,
and now have full data for 613 objects, significantly larger than the number (337) used in
earlier investigations.  The corresponding angular-size/redshift diagram gives
$\Omega_{\hbox{\scriptsize{m}}}=0.25+0.04/\!-0.03$, $\Omega_\Lambda=0.97+0.09/\!-0.13$
and $K=0.22+0.07/\!-0.10$ (68\% confidence limits).  In combination with supernova data,
and a simple-minded approach to CMB data based upon the angular size of the acoustic horizon,
our best figures are $\Omega_{\hbox{\scriptsize{m}}}=0.304+0.024/\!-0.023$,
$\Omega_\Lambda=0.693+0.034/\!-0.035$ and $K=-0.003+0.021/\!-0.019$.
We have examined a simple model of vacuum energy, based upon a scalar potential
$V(\phi)=\omega_{\hbox{\scriptsize{C}}}^2\phi^2/2$, to test the possibility that the vacuum
is dynamical; whereas the data favour a value $\omega_{\hbox{\scriptsize{C}}}=0$,
they are compatible with values in excess of the Hubble rate.   

\end{abstract}

\section{Introduction}

The angular-size/redshift relationship is in principle a simple cosmological test,
which has yet to play a noted part in the development of observational cosmology.
Endeavours in this field have generally concentrated upon classical double radio
sources as putative standard measuring rods, angular size being defined by core-lobe
or lobe-lobe separation.  Early results were problematical, showing angular sizes which
diminish more rapidly with increasing redshift $z$ than would be allowed by any plausible
cosmological model.  This behaviour is believed to be an evolutionary effect, brought
about by interaction with an evolving extra-galactic medium (Legg 1970; Miley 1971;
Barthel and Miley 1988; Singal 1988), or a selection effect, due to an inverse correlation
between linear size and radio power (Jackson 1973; Richter 1973; Masson 1980;
Nilsson \etal 1993).  There have been several attempts to disentangle the latter effects
from proper cosmological ones (Daly 1994; Buchalter \etal 1998).  Daly (1994) introduces
a model of lobe propagation which allows intrinsic size to be estimated from other source
parameters; Buchalter \etal (1998) select Fanaroff-Riley Type II sources (Fanaroff and
Riley 1974) with $z>0.3$ as the basis of their work, in part because their morphology
allows an unambiguous definition of angular size, and also allow for a linear-size/redshift
correlation of the sort discussed above; such measures bring the angular-size/redshift
diagram for extended sources into concordance with acceptable Friedmann cosmological models,
but not with sufficient precision to allow definitive statements about cosmological parameters.

More promising candidates in this context are ultra-compact radio sources, with milliarcsecond 
angular sizes measured by very-long-baseline interferometry (VLBI) (Kellermann 1993; Gurvits 1994).
Gurvits work was based upon a large VLBI 2.29 GHz survey undertaken by Preston \etal (1985)
(hereafter referred to as P85).  The latter was designed to establish 
a comprehensive full-sky list of ultra-compact radio sources; 1398 candidates
were selected from existing surveys, on the basis of total flux density at 2.7 GHz.
Selection criteria were complicated, but the candidate list is believed for example to 
be 97\% complete to a limit of 1.0 Jy for sources with spectral index $\alpha\geq -0.5$,
which comprise about 80\% of the list; further details are given in the original reference.
These sources were then observed in a series of VLBI experiments at 2.29 GHz; compact
structure was detected in 917 cases, and the corresponding correlated flux density $S_{\hbox{\scriptsize{c}}}$
(fringe amplitude) recorded, plus the total flux density $S_{\hbox{\scriptsize{t}}}$ if this was measured 
at the same time (531 out of the 917 cases).  As angular size is defined by fringe visibility 
$S_{\hbox{\scriptsize{c}}}/S_{\hbox{\scriptsize{t}}}$ (Thompson \etal 1986), the number of potential 
candidates for an investigation based upon P85 is thus appears to be 531.
In relation to these P85 give 269 redshifts; Gurvits (1994) added
further redshifts from V\'eron-Cetty and V\'eron (1991), to give a sample of 337
objects in all.  Gurvits gave plausible reasons for ignoring sources with $z<0.5$,
and using just the high-redshift data (258 objects) found marginal support for a
low-density Friedman cosmological model, but considered only those with vacuum energy
density parameter $\Omega_\Lambda=0$.  Using exactly the same data set (kindly supplied
by Dr. Gurvits), Jackson and Dodgson (1997, submitted 3/05/96) extended this work to
the full density-parameter plane, and concluded that the best values are $\Omega_{\hbox{\scriptsize{m}}}=0.2$
and $\Omega_\Lambda=0.8$, if the Universe is spatially flat, later refined to
$\Omega_{\hbox{\scriptsize{m}}}=0.24+0.09/-0.07$ (95\% confidence limits) (Jackson 2004).

The latter work develops an astrophysical model, according to  which the
underlying population consists of compact symmetric objects (Wilkinson \etal 1994),
comprising central low-luminosity cores straddled by two mini-lobes.
Ultra-compact objects are identified as cases in which the lobes are moving
relativistically and are close to the line of sight, when D\H oppler boosting
allows just that component which is moving towards the observer to be observed;
the interferometric angular sizes upon which this work is based correspond to
the said components.  As $z$ increases a larger D\H oppler factor is required;
it turns out that latter approximately cancels the cosmological redshift, so 
that the observed component is seen in its rest frame.  This model, coupled
with the fact that the central engines which power these sources are reasonably
standard objects (black holes with masses close to $1.5\times 10^{10}M_\odot$),
gives a plausible account of their behaviour as standard measuring rods, and
of why those with $z<0.5$ should ignored.  Those having $z\geq 0.5$ are the
ultra-compact analogues (and perhaps the precursors) of Fanaroff-Riley Type II sources, 
and there are parallels between this work and that of Buchalter \etal (1998).   
This is a slightly oversimplified precis of the account given in Jackson (2004),
to which the reader is referred for further details.

We have taken a fresh look at the P85 catalogue, and updated the latter with
respect to both redshift and radio information.  Details of the new data set
are given in Section 2.  Section 2 also discusses the acoustic horizon and its
linear size; coupled with accurate measures of the location of the first 
peak in angular spectrum of the cosmic microwave background (CMB) radiation
(Balbi \etal 2000; de Bernardis \etal 2000; Hanany \etal 2000, Hinshaw \etal 2003),
these allow the WMAP results to be interpreted in a simple and transparent fashion,
which does not need the full machinery of CMBFAST (Seljak and Zaldarriaga 1996),
that is if particular figures for Hubble's constant and baryon content are accepted.
Cosmological results are presented in Section 3, for the new data alone and in
combination with the most recent supernova observations (Riess \etal 2004;
Astier \etal 2006), mainly as marginalized confidence regions in the
$\Omega_{\hbox{\scriptsize{m}}}$--$\Omega_\Lambda$ plane.  Section 4 considers
the evolution of vacuum energy in the light of the new data.

Although we believe that the results to be presented here are of intrinsic interest, 
a large part of our motivation in undertaking this work is to enhance further the 
credibility of milliarcsecond radio-sources as standard measuring rods.  In this context
it should be noted that their first application to the $\Omega_{\hbox{\scriptsize{m}}}$--$\Omega_\Lambda$
problem (Jackson and Dodgson 1997) was blind, in the sense that it gave answers close
to the concordance model at a time when Type Ia supernovae were giving $\Omega_{\hbox{\scriptsize{m}}}\sim 1$
(Perlmutter \etal 1997).  This situation did not change until the advent of
Schmidt \etal (1998), Riess \etal (1998) and Perlmutter \etal (1999).

\section{Data}

\begin{figure}[here]
\begin{center}
\includegraphics[width=12.5cm] {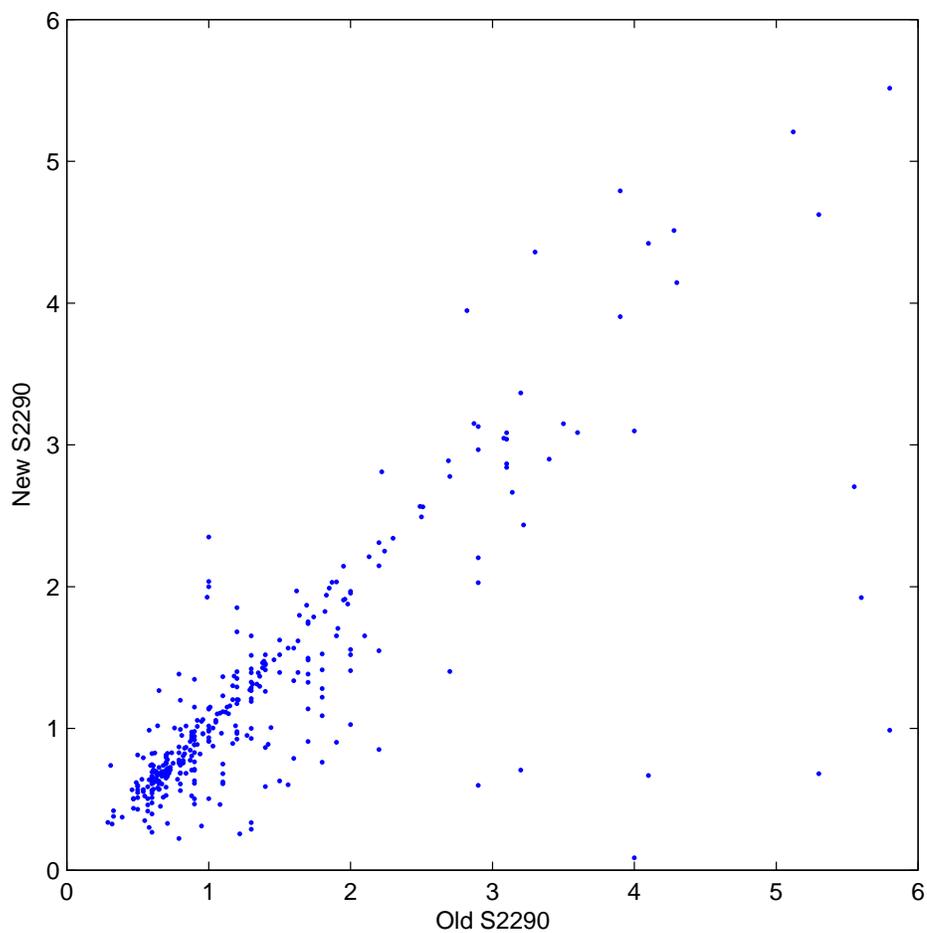}
\end{center}
\caption{\label{FigS} Legacy data.  Plot of flux density S2290 derived from the Parkes catalogue
PKS90 (New S2290) against the corresponding figure from Preston \etal (1985) (Old S2290),
for those sources (346) which have measures in both. Values of S2290 derived from PKS90
are used in suitable cases when a direct measure is not given in Preston \etal (1985).}  
\end{figure}

Of the 1398 sources in the P85 catalogue, 917 have definite measures of correlated flux 
density (the rest have upper limits), and 531 have both total and correlated flux densities.
We have consulted the NASA/IPAC Extragalactic Database (NED), and find 957, 723 and 456
redshifts for the respective sub-samples (end of August 2005).  Thus we have redshifts and
visibilities for 456 sources; formally 8 of these have visibility$>$1 and were discarded;
the remaining 448 comprise our gold standard sub-sample.  There are 386 sources for which a
correlated flux density is listed but no total flux density (531+386=917); we have attempted
to update this sub-sample, by searching for convenient listings in the literature which are roughly
contemporaneous with P85 and at a frequency not too far removed from 2.29 GHz.  With respect to
southern hemisphere sources, we find that the Parkes catalogue PKSCAT90 (Parkes Catalogue 1990)
serves this purpose very well.  The latter gives total flux densities at 2.7 Ghz, which we have
extrapolated to 2.29 GHz using the spectral indices given in P85 (in the few cases where $\alpha$
is not listed we have used a value of zero).  This procedure gives 256 new values of $S_{\hbox{\scriptsize{t}}}$,
for which we have 174 redshifts; formally 9 of these have visibility$>$1 and were discarded;
the remaining 165 sources comprise our silver standard sub-sample, which we expect to be of
lower quality than the gold.  We have checked the validity of this approach by considering
the overlap: Figure \ref{FigS} is a plot of the total flux density so extrapolated against
that listed in P85, for all sources (346) for which we have both; the two figures are well
correlated, with a correlation coefficient of 0.854.  With respect to northern hemisphere sources,
we have examined the NRAO/VLA Sky Survey (NVSS) (Condon \etal 1998), which gives $S_{\hbox{\scriptsize{t}}}$
at 1.4 GHz, but find that this data set (our bronze sub-sample) is less useful, being too far removed in
frequency and epoch.  The sample used here will comprise our gold plus silver sub-samples,
amounting to 613 objects out of a possible 917, with a range $z=0.0035$ to $z=3.787$. 

Preston \etal (1985) selected their list of 1398 VLBI candidates mainly from an earlier version
of the Parkes survey (PKSCAT85, see Parkes Catalogue 1990), and from the NRAO-Bonn survey
(K\H uhr \etal 1981); it would have been useful for this investigation if the corresponding
flux densities had been included in P85, but this was not the case.  Almost all of those
sources in K\H uhr \etal (1981) were re-measured at 2290 GHz by Preston \etal (1985).
  
Regarding the acoustic horizon, there are several publications which give similar analytical
formulae for its linear size $l_{\hbox{\scriptsize{AcH}}}$ (Hu and Sugiyama 1995; Mukhanov 2004).
Writing that given by Hu and Sugiyama (1995) in terms of quantities which are all evaluated at recombination
of the photon-baryon plasma, we find 

\begin{equation}\label{A}
l_{\hbox{\scriptsize{AcH}}}={1 \over \sqrt{3}}{C \over A}\ln\left({A\sqrt{1+X_{\hbox{\scriptsize{R}}}}+\sqrt{A^2+1}
\over A\sqrt{X_{\hbox{\scriptsize{R}}}}+1}\right)
\end{equation}

\noindent where

\begin{equation}\label{B}
A=\left({3 \over 4}{\rho_{\hbox{\scriptsize{b}}} \over \rho_\gamma}\right)^{1/2}_{\hbox{\scriptsize{r}}}
=\left({3 \over 4}{\rho_{\hbox{\scriptsize{c}}} \over \rho_\gamma}
{\Omega_{\hbox{\scriptsize{b}}} h^2 \over 1+z_{\hbox{\scriptsize{r}}}}\right)^{1/2}_0
\end{equation}

\begin{equation}\label{C}
X_{\hbox{\scriptsize{R}}}=\left({\rho_{\hbox{\scriptsize{m}}} \over \rho_{\hbox{\scriptsize{R}}}}\right)
^{-1}_{\hbox{\scriptsize{r}}}=\left({\rho_{\hbox{\scriptsize{c}}} \over \rho_{\hbox{\scriptsize{R}}}}
{\Omega_{\hbox{\scriptsize{m}}} h^2 \over 1+z_{\hbox{\scriptsize{r}}}}\right)^{-1}_0
\end{equation}

\begin{equation}\label{D}
C=2c\left({8\pi G\rho_{\hbox{\scriptsize{m}}} \over 3}\right)^{-1/2}_{\hbox{\scriptsize{r}}}
=2c(\Omega_{\hbox{\scriptsize{m}}} H_0^2)^{-1/2}(1+z_{\hbox{\scriptsize{r}}})^{-3/2}.
\end{equation}
\vskip 0.3cm

\noindent Here $\rho_{\hbox{\scriptsize{c}}}$ is the critical density corresponding to a particular value
of Hubble's constant $H_0=100h$ km sec$^{-1}$ Mpc$^{-1}$; subscripts b and m
refer to baryons and baryons+cold dark matter, and $\gamma$ and R to CMB photons
and photons+neutrinos, respectively; a subscript r refers to values at recombination,
and zero to current values.  Equation (\ref{A}) is exact only in the case of flat cosmologies,
but in practice flatness will hold to a high degree of accuracy for any realistic model,
over the interval between the big bang and recombination.  It is exact also only in the
case of instantaneous recombination at some redshift $z_{\hbox{\scriptsize{r}}}$, which is unrealistic;
in reality recombination starts at a redshift of about 1200 and is complete at about 800.
For this reason equation (\ref{A}) underestimates the size of the acoustic horizon,
because it does not allow for those charged baryons which recombine in the period
immediately preceding $z_{\hbox{\scriptsize{r}}}$, and curtails acoustic propagation too early (Mukhanov 2004).
We have allowed for this by devising a semi-empirical two-parameter fitting formula,
which increases $l_{\hbox{\scriptsize{AcH}}}$ by a factor $\alpha(\Omega_{\hbox{\scriptsize{b}}},h)$, and reduces
$\rho_{\hbox{\scriptsize{R}}}$ by a fixed factor $\beta$; the latter is effective because it allows more time for acoustic propagation to take place.  Taking $z_{\hbox{\scriptsize{r}}}=1088$ (Hinshaw \etal 2003) and three massless neutrinos, we find that the following values are appropriate
\footnote[3]{Including $z_{\hbox{\scriptsize{r}}}$ dependence the equation is:
$\alpha=1.1991+7.0\times 10^{-4}(z_{\hbox{\scriptsize{r}}}-1080)\newline
+[1.020-1.5\times 10^{-3}(z_{\hbox{\scriptsize{r}}}-1080)](\Omega_{\hbox{\scriptsize{b}}}-0.04)
+[0.121-4.0\times 10^{-4}(z_{\hbox{\scriptsize{r}}}-1080)](h-0.7)$}
:

\begin{equation}\label{E}
\alpha=1.2047+1.008(\Omega_{\hbox{\scriptsize{b}}}-0.04)+0.1178(h-0.7)
\end{equation}

\begin{equation}\label{F}
\beta=0.602.
\end{equation}
\vskip 0.3cm

\begin{figure}[here]
\begin{center}
\includegraphics[width=12.5cm] {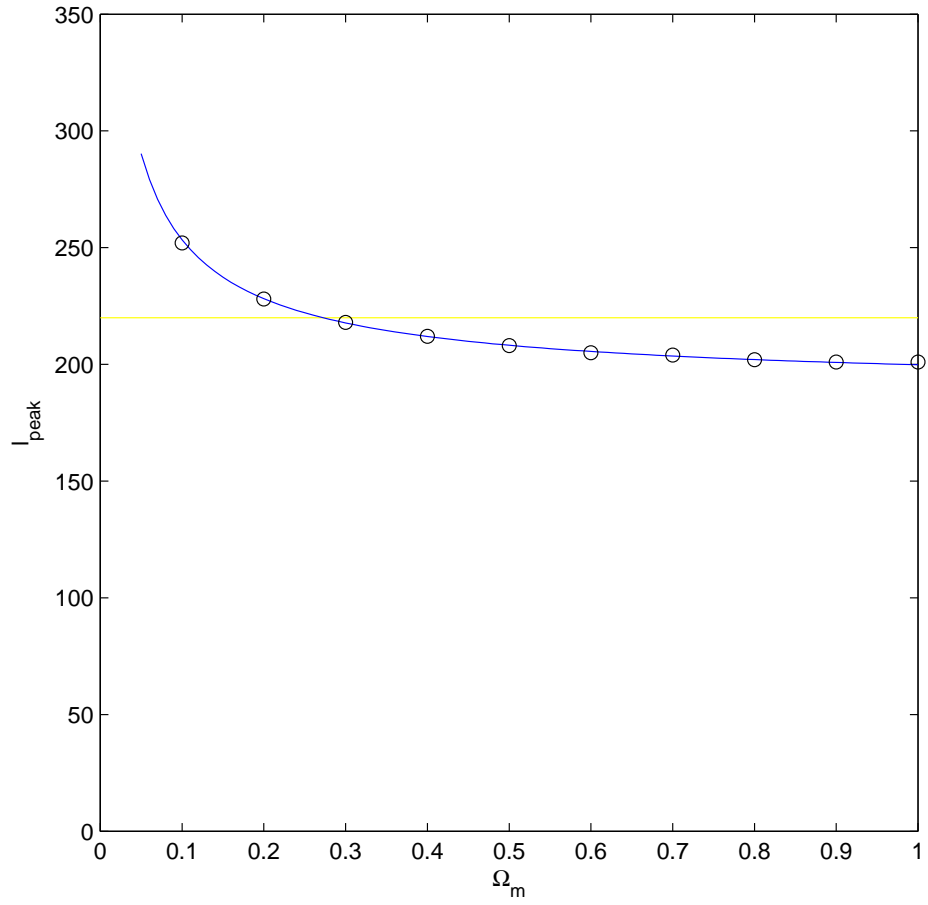}
\end{center}
\caption{\label{FigAcH} Multipole moment $l_{\hbox{\scriptsize{peak}}}$ corresponding to the first acoustic
peak in the CMB angular spectrum; blue line is the analytical curve according to
equations (\ref{A}), (\ref{E}) and (\ref{F}), with $z_{\hbox{\scriptsize{r}}}=1088$, $h=0.7$ and $\Omega_{\hbox{\scriptsize{b}}}=0.04$;
data points are from CMBFAST; all spatially flat: $\Omega_\lambda=1-\Omega_{\hbox{\scriptsize{m}}}$.
The yellow horizontal line is the WMAP figure $l_{\hbox{\scriptsize{peak}}}=220$.}  
\end{figure}

\noindent Comparison with numerical results from CMBFAST show that with these modifications equation (\ref{A}) 
is accurate to better than 0.4\% over the ranges $0.03\leq \Omega_{\hbox{\scriptsize{b}}}\leq 0.05$,
$0.65\leq h\leq 0.75$ and $0.2\leq \Omega_{\hbox{\scriptsize{m}}}\leq 0.5$, and to better than 0.6\% if the latter
range is increased to $0.2\leq \Omega_{\hbox{\scriptsize{m}}}\leq 1.0$.  We convert $l_{\hbox{\scriptsize{AcH}}}$
into an angular scale using the correct angular-diameter distance $d_{\hbox{\scriptsize{A}}}(z)$, which need not correspond to a spatially flat model.  As an example, Figure \ref{FigAcH} shows the multipole index
$l_{\hbox{\scriptsize{peak}}}$ corresponding to the first D\H oppler peak in the CMB angular spectrum as
a function of $\Omega_{\hbox{\scriptsize{m}}}$, according to equation (\ref{A}), in the case of spatial flatness,
$\Omega_\Lambda=1-\Omega_{\hbox{\scriptsize{m}}}$.

For Type Ia supernovae (SNe Ia) we have used a composite data set comprising objects from 
Riess \etal (2004) (the gold sample) and Astier \etal (2006) (The Supernova Legacy Survey, SNLS).
Each of these samples includes a low-$z$ subset ($z\leq 0.125$, 68 and 44 objects respectively);
the two low-$z$ subsets are not identical, but have 21 objects in common.  Our composite data set
comprises the full gold sample from Riess \etal (2004) (157 objects) plus 71 high-z objects from
Astier \etal (2006) (Table 9 of their paper); to make sure that no objects were counted twice,
we do not include the SNLS low-$z$ subset in our composite sample.  However, before assembling 
the latter there is a matching problem to be considered, discussed below.  

The distance indicator listed in both Riess \etal (2004) and Astier \etal (2006) is distance
modulus $\mu_B$; to bring the two samples into conformity we find that it is necessary reduce the 
gold subset distance moduli by 0.14 magnitudes.  This figure was determined by fitting the equation

\begin{equation}\label{G}
\mu_B=5\log_{10}(z)+C
\end{equation}

\noindent first to the low-$z$ gold subset, and then to the low-$z$ SNLS one; the 
best values are $C=43.37$ and $C=43.23$ respectively.  We choose the low-$z$ subsets
for this purpose (rather than the full ones), because they overlap, and because the
outcome does not depend upon a choice of cosmological parameters.  

The  probable reasons for this offset are outlined below.  Astier \etal (2006) list a 
rest-frame maximum-luminosity magnitude $m_B^*$ for each supernova, from which they derive
a corresponding magnitude $m_B$ corrected for brighter-slower and brighter-bluer correlations,
which reduce the dispersion in magnitude by a factor of about 2 (see for example Guy \etal 2005
and the references given there).  The magnitudes so corrected were used in a Hubble diagram
to derive a best-fitting absolute magnitude $M=-19.31+5\log_{10}(h/0.7)$, which was subtracted
from the corrected magnitudes to give the listed distance moduli.  The latter thus depend on
the choice of Hubble's constant; Astier \etal (2006) adopt $h=0.7$ as their fiducial value.
Although knowledge of these parameters is not essential to the determination of cosmological
parameters, which depend only upon magnitude differences, they are nevertheless useful adjuncts.  
Riess \etal (2004) follow a similar procedure, but do not mention values of $H_0$ and $M$
explicitly (see the discussion in their Appendix A); there is no reason to suppose that the
implict values are the same as those used by Astier \etal (2006).  Having reduced the gold
values by 0.14 magnitudes before combining the two lists, it is a matter of taste whether
we use distance moduli or corrected magnitudes as working variables; we prefer the latter,
and have reconstructed a table of same by adding $-19.31$ to each $\mu_B$ in our composite list.
In what follows we marginalize over the absolute magnitude $M$, rather than treat it as a fixed
parameter.

\section{Cosmological parameters}

With respect to the milliarcsecond radio-sources, our procedure follows that given in Jackson (2004).
Of the 613 sources mentioned in Section 2, 468 have $z>0.5$, with $0.501\leq z\leq 3.787$.  These
are placed in appropriate redshift bins, and the mean redshift and a suitably defined characteristic
angular size $\theta$ relating to each bin comprise our data points; our working numbers are logs
of $\theta$.  As in Jackson (2004), we find that the best compromise (between many bins containing
few objects and a small number containing many) is to have 6  bins, which conveniently is a divisor
of 468 to give 78 objects in each.  This number of bins is comfortably larger than the number of
parameters to be determined, and the correspondingly large number of objects within in each accurately
determines the characteristic angular size.  To derive confidence regions, we give each point equal
weight, and define a corresponding standard deviation: $\sigma^2=\hbox{residual sum-of-squares/}(n-p)$,
relative to the best-fitting curve defined by $\Omega_{\hbox{\scriptsize{m}}}$, $\Omega_\Lambda$ and $d$,
where $n=6$ is the number of points and $p=3$ is the number of fitted parameters; here $d$ a characteristic
linear size associated with the source population.  The definition of `characteristic' in this context
is discussed at length in Jackson (2004), where it is shown that there is a clear inverse correlation
between linear size and absolute radio luminosity.  In a flux-limited sample those sources with large 
redshift are intrinsically the most powerful, which means that the corresponding size distribution is 
biased towards intrinsically smaller objects; this introduces a weak bias towards open universes, if
characteristic angular size within each bin is taken to be the simple mean.  As a countermeasure the
characteristic angular size is defined as a median biased towards the lower end of the size distribution,
namely the boundary between the bottom third and the top two thirds of the latter.
With 78 objects in each bin the boundary is defined by the 26th point; in fact we take the mean of points
16 to 36 (i.e. $26 \pm 10$) as a somewhat smoother measure than the 26th point alone, but this range
is not critical.  With this definition we find best-fitting values $\Omega_{\hbox{\scriptsize{m}}}=0.22$, $\Omega_\Lambda=1.06$ and $d=7.75h^{-1}$ kpc, and a standard deviation $\sigma=0.0074$ in $\log\theta$,
i.e. 1.7\% in $\theta$.  To show that these figures are robust with respect to binning, we have repeated the calculation for 15 bin sizes, from 13 bins of 36 objects to 6 bins of 78 objects, in steps of 3 objects,
retaining the maximum redshift of 3.787 in each case and using the largest number of bins compatible with
having no objects with $z<0.5$.  The outcome is $\Omega_{\hbox{\scriptsize{m}}}=0.24\pm 0.04$,
$\Omega_\Lambda=1.01\pm 0.07$, and curvature parameter $K=0.25\pm 0.04$ (68\% confidence limits).
These figures are consistent with those derived from a confidence region, discussed below.
The question of robustness with respect to the definition of characteristic angular size will be
considered at the end of this section. 

\begin{figure}[here]
\begin{center}
\includegraphics[width=12.5cm] {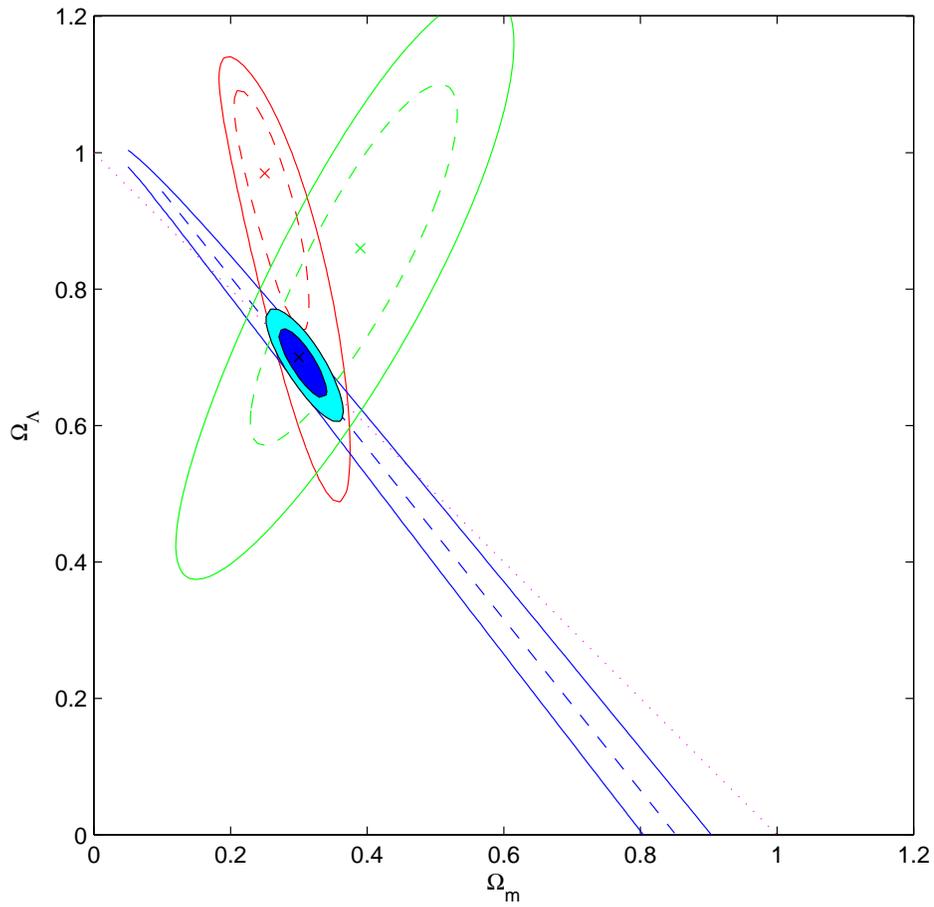}
\end{center}
\caption{\label{FigOmega_sigmas} Confidence regions in the $\Omega_{\hbox{\scriptsize{m}}}$--$\Omega_\Lambda$ plane, 95\% and 68\%; red=milliarcsecond radio-sources, blue=acoustic horizon, green=Type Ia supernovae; the filled 
areas are joint confidence regions for all three; the acoustic horizon fit is exact along 
the dashed blue line.} 
\end{figure}

Rather than rely too heavily upon a particular binning, we have produced points corresponding to 
bin sizes of 76, 77 and 78 objects, for which the respective standard deviations in $\log\theta$
are $\sigma=0.0125$, $0.0031$ and $0.0074$; our working points are a composite of these three
cases.  We do not add smaller bins, because this generally means discarding valuable data close
to $z=0.5$.  The resulting data points are shown in Table \ref{Table1}; the standard deviation is now
$\sigma=0.00603$.  The red plots in Figure \ref{FigOmega_sigmas} shows 68\% and 95\% confidence regions in the
$\Omega_{\hbox{\scriptsize{m}}}$--$\Omega_\Lambda$ plane, marginalized over the nuisance parameter $d$;
1-dimensional projections give $\Omega_{\hbox{\scriptsize{m}}}=0.25+0.04/-0.03$, $\Omega_\Lambda=0.97+0.09/-0.13$
and $K=0.22+0.07/-0.10$ (68\% confidence limits).

\begin{table}[here]
\caption{\label{Table1}Data points for the angular-size/redshift relationship;
$\theta$ is in milliarcseconds.}
\begin{indented}
\item[]\begin{tabular}{@{}cc}
\br
$z$ & $\theta$ \\
\mr
0.6153 & 1.4624 \\
0.8580 & 1.2801 \\
1.1527 & 1.1599 \\
1.4200 & 1.1448 \\
1.8288 & 1.1760 \\
2.5923 & 1.2374 \\
\br
\end{tabular}
\end{indented}
\end{table}

With respect to the acoustic horizon, the key observational point is the position of the first peak in the 
CMB angular spectrum, now located precisely at multipole moment $l_{\hbox{\scriptsize{peak}}}=220.1\pm 0.8$
(Hinshaw \etal 2003).  We must first specify values of $H_0$ and $\Omega_{\hbox{\scriptsize{b}}} h^2$,
and will adopt the Wilkinson Anisotropy Microwave Probe (WMAP) values $H_0=72\pm 5$ km sec$^{-1}$ Mpc$^{-1}$
and $\Omega_{\hbox{\scriptsize{b}}} h^2=0.024\pm 0.001$ (68\% confidence limits) (Spergel \etal 2003).  Strictly speaking we should not use values which derive from the angular spectrum, but these figures are well supported
by independent observations, Cepheid variables in the case of $H_0$ (Freedman \etal 2001; Altavilla \etal 2004;
Riess \etal 2005), and nucleosynthesis considerations for $\Omega_{\hbox{\scriptsize{b}}} h^2$ (D'Oderico \etal 2001;
Kirkman \etal 2003; Pettini and Bowen 2001).  If equation (\ref{A}) is used to calculate the size of the acoustic horizon the biggest source of uncertainty arises from the uncertainty in Hubble's constant;
$l_{\hbox{\scriptsize{AcH}}}$ changes by $\pm 2.9$\% as $H_0$ changes over the above error range,
whereas the corresponding figure for $\Omega_{\hbox{\scriptsize{b}}} h^2$ is $\pm 0.28$\%.  
Thus our procedure is to fix $\Omega_{\hbox{\scriptsize{b}}} h^2$ at $0.024$, set
$l_{\hbox{\scriptsize{AcH}}}=d_{\hbox{\scriptsize{A}}}(z_{\hbox{\scriptsize{r}}})\times \pi/220$, and regard
equation (\ref{A}) as giving $h$ as a function of $\Omega_{\hbox{\scriptsize{m}}}$ and $\Omega_\Lambda$. 
Equation (\ref{A}) is solved to give $h$ at each point in the $\Omega_{\hbox{\scriptsize{m}}}$--$\Omega_\Lambda$
plane, where $d_{\hbox{\scriptsize{A}}}(z_{\hbox{\scriptsize{r}}})$ is the corresponding angular-diameter distance;
we then define values $\chi^2=[(h-0.72)/0.05]^2$, which with 1 degree of freedom gives the appropriate confidence
region.  Note that the error in $l_{\hbox{\scriptsize{peak}}}$ is quite negligible in this context.
The continuous blue lines in Figure \ref{FigOmega_sigmas} define 95\% confidence limits; the fit is exact 
along the dashed blue line, which has slope $-1.255$ and intersects the flat (magenta dotted) line at 
$\Omega_{\hbox{\scriptsize{m}}}=0.271$.  The slight curvature in the contours at the low-density end of the
diagram is because radiation still makes a significant contribution to the total density at
$z_{\hbox{\scriptsize{r}}}$: $(\rho_{\hbox{\scriptsize{R}}}/\rho_{\hbox{\scriptsize{m}}})_{z_{\hbox{\scriptsize{r}}}}=0.45$ when $\Omega_{\hbox{\scriptsize{m}}}=0.1$.

The green plots in Figure \ref{FigOmega_sigmas} show 68\% and 95\% confidence regions for $157+71=228$
SNe Ia, marginalized over absolute magnitude $M$.  In weighting each point we have followed exactly
the scheme given in Astier \etal (2006); each magnitude is assigned a variance $\sigma_p^2+\sigma_{int}^2$,
where $\sigma_p$ is the listed photometric error, and $\sigma_{int}$ is a fixed error attributed to intrinsic
variations, which is chosen to give a minimum value of $\chi^2$ equal to the expected figure $225$.
We find $\sigma_{int}=0.105$ for our composite sample; we have checked the corresponding figure
for the full (nearby plus distant) SNLS sample, and find $\sigma_{int}=0.131$, which coincides
with the value given in Astier \etal (2006).

A summary of best-fitting cosmological parameters is given in Tables \ref{Table295} and \ref{Table268},
for the three basic samples separately and in various combinations.

\begin{table}[here]
\caption{\label{Table295} Best-fitting cosmological parameters, 95\% confidence limits;
MAS=milliarcsecond radio-sources, AcH=acoustic horizon, SN=Type Ia supernovae.}
\begin{indented}
\item[]\begin{tabular}{@{}cccc}
\br
Sample & $\Omega_{\hbox{\scriptsize{m}}}$ & $\Omega_\Lambda$ & K\\
\mr
MAS &        $0.25 +0.09/\!-0.06 $&$ 0.97 +0.15/\!-0.34 $&$  0.22 +0.11/\!-0.26$ \\
MAS+AcH &    $0.31 +0.08/\!-0.06 $&$ 0.69 +0.09/\!-0.11 $&$ -0.01 +0.04/\!-0.05$ \\
SN &         $0.39 +0.18/\!-0.21 $&$ 0.86 +0.31/\!-0.38 $&$  0.25 +0.47/\!-0.57$ \\
SN+AcH &     $0.30 +0.07/\!-0.06 $&$ 0.69 +0.08/\!-0.09 $&$ -0.01 +0.04/\!-0.04$ \\  
MAS+SN &     $0.31 +0.05/\!-0.04 $&$ 0.76 +0.12/\!-0.15 $&$  0.07 +0.12/\!-0.14$ \\  
MAS+SN+AcH & $0.30 +0.05/\!-0.04 $&$ 0.69 +0.06/\!-0.07 $&$  0.00 +0.04/\!-0.04$ \\
\br
\end{tabular}
\end{indented}
\end{table}

\begin{table}[here]
\caption{\label{Table268} Best-fitting cosmological parameters, 68\% confidence limits;
MAS=milliarcsecond radio-sources, AcH=acoustic horizon, SN=Type Ia supernovae.}
\begin{indented}
\item[]\begin{tabular}{@{}cccc}
\br
Sample & $\Omega_{\hbox{\scriptsize{m}}}$ & $\Omega_\Lambda$ & K\\
\mr
MAS &        $0.249 +0.041/\!-0.031 $&$ 0.969 +0.089/\!-0.135 $&$  0.218 +0.066/\!-0.103$ \\
MAS+AcH &    $0.308 +0.038/\!-0.033 $&$ 0.687 +0.047/\!-0.054 $&$ -0.005 +0.022/\!-0.024$ \\
SN &         $0.389 +0.096/\!-0.104 $&$ 0.856 +0.165/\!-0.183 $&$  0.245 +0.249/\!-0.277$ \\
SN+AcH &     $0.300 +0.033/\!-0.032 $&$ 0.694 +0.040/\!-0.042 $&$ -0.006 +0.020/\!-0.020$ \\  
MAS+SN &     $0.309 +0.024/\!-0.024 $&$ 0.763 +0.065/\!-0.071 $&$  0.072 +0.063/\!-0.070$ \\  
MAS+SN+AcH & $0.304 +0.024/\!-0.023 $&$ 0.693 +0.034/\!-0.035 $&$ -0.003 +0.021/\!-0.019$ \\ 
\br
\end{tabular}
\end{indented}
\end{table}

\begin{figure}[here]
\begin{center}
\includegraphics[width=12.5cm] {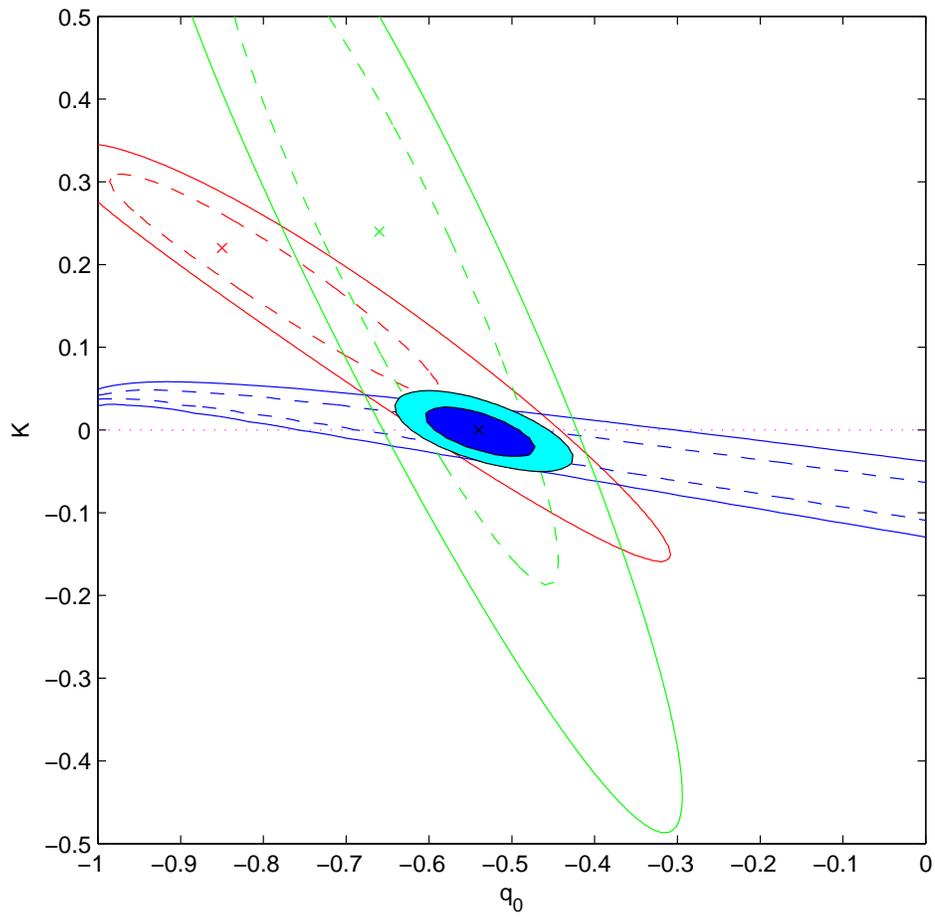}
\end{center}
\caption{\label{Figq0_sigmas} Confidence regions in the $q_0$--$K$ plane, 95\% and 68\%;
red=milliarcsecond radio-sources, blue=acoustic horizon, green=Type Ia supernovae;
the filled areas are joint confidence regions for all three;
the dashed blue lines are now 68\% confidence limits.}  
\end{figure}

Figure \ref{Figq0_sigmas} is an alternative presentation of the same information, in which the coordinates have been transformed to the deceleration parameter $q_0=\Omega_{\hbox{\scriptsize{m}}}/2-\Omega_\Lambda$ and curvature
parameter $K=\Omega_{\hbox{\scriptsize{m}}}-\Omega_\Lambda-1$.  The best figures for $q_0$ are $-0.55\pm 0.09$
(95\%) and $\pm 0.05$ (68\%).  Figure \ref{Figq0_sigmas} is clearly compatible with the conclusion that we are living
in a spatially flat Universe with accelerating expansion.

The outstanding question relates to robustness of the MAS figures with respect to definition
of characteristic angular size.  If we use simple means instead of biased medians, 68\% figures
for MAS alone become $\Omega_{\hbox{\scriptsize{m}}}=0.28+0.06/-0.04$, $\Omega_\Lambda=0.74+0.18/-0.30$
and $K=0.02+0.14/-0.25$, which represent a significant shift towards open universes, but not to an
extent which alters the overall qualitative picture.  This shift illustrates the point made in
the first paragraph of this section, and is precisely the effect found in Jackson (2004).
With respect to composite data sets the shift is of virtually no consequence, figures for
MAS+SN+AcH becoming $\Omega_{\hbox{\scriptsize{m}}}=0.29+0.03/-0.02$, $\Omega_\Lambda=0.70+0.03/-0.04$
and $K=-0.01+0.02/-0.02$.

\section{Evolution}

We do not attempt here to map out the evolution of vacuum energy with cosmic
time $t$.  The concensus is that large volumes of high-quality data will be
needed to do this in an unambiguous way (Aldering 2005; Mosoni \etal 2006),
an order of magnitude greater than those available currently.  The vacuum
equation-of-state is characterised by the ratio of vacuum pressure to vacuum
energy density, denoted by $w(z)$.  Simple functional forms have been used to model
evolution, for example $w=w_0+w_1z$ (Cooray and Huterer 1999) and $w=w_0+w_1z/(1+z)$
(Linder 2003); however, Maor \etal (2002) have shown by means of simulated data and
counter-example that such parametrizations can produce very misleading results.
If $w$ is assumed to be constant then the best-fitting value does not necessarily
coincide with the time-averaged value.  In general there is a propensity to produce
apparently well-fitting functions $w(z)$ which are distinctly more negative than they should be,
in some cases significantly less than $-1$; the latter are generally regarded as unphysical
(see also Maor \etal 2001; Linder 2004; J\H onsson \etal 2004; Bassett \etal 2004,
Jassal \etal 2006).  There are also non-parametric approaches (Chiba and Nakamura 2000;
Daly and Djorgovski 2003, 2004; Gerke and Efstathiou 2002; Huterer and Turner 1999, 2001;
Huterer and Starkman 2003; Wang and Freese 2006)

As an alternative to the simple functional forms $w(z)$ mentioned above, we shall use
a proper physical model of dynamical vacuum energy, based upon the simple potential

\begin{equation}\label{V}
V(\phi)=\omega_{\hbox{\scriptsize{C}}}^2\phi^2/2
\end{equation}

\noindent where $\omega_{\hbox{\scriptsize{C}}}$ is the Compton frequency of the corresponding boson.
If it turns out that the data favour a value $\omega_{\hbox{\scriptsize{C}}}>0$,  
we shall take this as evidence that the Universe is governed by a scalar vacuum energy
which is diminishing with time in a reasonably smooth way, rather than a cosmological constant,
and not as support for the particular potential (\ref{V}).  In other words we are using the
latter to test the hypothesis that the rate-of-change of vacuum energy is zero.  
In what follows we shall assume that the Universe is spatially flat, and that the true
cosmological constant is strictly zero.  Corresponding to the said potential, we have scalar density     

\begin{equation}\label{H} 
\rho_\phi=(\dot\phi^2+\omega_{\hbox{\scriptsize{C}}}^2\phi^2)/2
\end{equation}

\noindent and pressure

\begin{equation}\label{I}
p_\phi=(\dot\phi^2-\omega_{\hbox{\scriptsize{C}}}^2\phi^2)/2.
\end{equation}

\noindent The relevant Friedmann equation for the scale factor $R(t)$ is

\begin{equation}\label{J}
\ddot R=-{4\pi \over 3}G(\rho_\phi+3p_\phi+\rho_{\hbox{\scriptsize{m}}}+2\rho_{\hbox{\scriptsize{R}}})R
=-{4\pi \over 3}G(2\dot\phi^2-\omega_{\hbox{\scriptsize{C}}}^2\phi^2+\rho_{\hbox{\scriptsize{m}}}
+2\rho_{\hbox{\scriptsize{R}}})R~~~~~~
\end{equation}

\noindent and the scalar field is governed by the equation

\begin{equation}\label{K}
\ddot\phi+3H\dot\phi+\omega_{\hbox{\scriptsize{C}}}^2\phi=0
\end{equation}

\noindent where $H=\dot R/R$ is the instantaneous value of Hubble's `constant'. 

To solve the system of equations (\ref{J}) to (\ref{K}), it is convenient to introduce
dimensionless variables $x=R/R_0$, $y=\dot x=\dot R/R_0$, and

\begin{equation}\label{L}
u=\left({4\pi G \over 3}\right)^{1/2}\phi~~~~~~~~~~~~v=\dot u=\left({4\pi G \over 3}\right)^{1/2}\dot\phi
\end{equation}

\noindent where the dot is now the derivative with respect $H_0t$, in other words our unit of time
is the Hubble time.  Equations (\ref{J}) to (\ref{K}) are then re-cast in the form
of an autonomous non-linear dynamical system:
\bigbreak

\begin{equation}\label{EQ}
\end{equation}
\vskip -1.2cm
$$
\eqalign{
\dot u&=v~~~~~~~~~~~~\dot v=-3yv/x-\omega_{\hbox{\scriptsize{C}}}^2u\cr
\dot x&=y~~~~~~~~~~~~\dot y=-(2v^2-\omega_{\hbox{\scriptsize{C}}}^2u^2)
-\textstyle{1 \over 2}\Omega_{\hbox{\scriptsize{m}}}/x^2-\Omega_{\hbox{\scriptsize{R}}}/x^3.\cr
}
$$

\noindent Here $\omega_{\hbox{\scriptsize{C}}}$ is $\omega_{\hbox{\scriptsize{C}}}/H_0$ in the old units
(rather than introduce a new symbol we trust that context will dictate the appropriate form) and
$\Omega_{\hbox{\scriptsize{R}}}=8.023\times 10^{-5}$, corresponding to $h=0.72$.  In terms of the new
variables the angular-diameter distance is

\begin{equation}\label{DA}
d_{\hbox{\scriptsize{A}}}(z)=(1+z)^{-1}{c \over H_0}\int_{t(z)}^{t_0}{dt \over x(t)}\,.
\end{equation} 

By definition initial conditions (at $t=t_0$) are $x(t_0)=1$ and $y(t_0)=1$; those relating to
$u$ and $v$ are in part determined in part by a specified value of the scalar density parameter
$\Omega_\phi=8\pi G\rho_\phi(t_0)/3H_0^2$, which using equations (\ref{H}) and (\ref{L}) becomes

\begin{equation}\label{M}
\Omega_\phi=\omega_{\hbox{\scriptsize{C}}}^2u(t_0)^2+v(t_0)^2.
\end{equation}

\noindent However, equation (\ref{K}) is second order, and we must specify initial values $u(t_0)$
and $v(t_0)=\dot u(t_0)$ to give a unique solution, or equivalently current values of the scalar density
{\it and} pressure.  We need to find an extra condition which fixes $u(t_0)$ and $v(t_0)$ unambiguously.
To this end, we note that when $z~^>_\sim~3$ we expect matter or radiation to be dominant, and
$R(t)\propto t^n$, with $n=2/3$ or $1/2$ respectively; equation (\ref{K}) then becomes

\begin{equation}\label{N}
\ddot\phi+{3n \over t}\dot\phi+\omega_{\hbox{\scriptsize{C}}}^2\phi=0.
\end{equation}

\noindent Equation (\ref{N}) has a regular singular point at $t=0$, and has a two-parameter series
solution, for example

\begin{equation}\label{N2}
\phi=A[1-(\omega_{\hbox{\scriptsize{C}}}t)^2/6+...]+{B \over t}[1-(\omega_{\hbox{\scriptsize{C}}}t)^2/2+...]
\end{equation}

\noindent when $n=2/3$, and

\begin{equation}\label{N3}
\phi=A[1-(\omega_{\hbox{\scriptsize{C}}}t)^2/5+...]+{B \over t^{1/2}}[1-(\omega_{\hbox{\scriptsize{C}}}t)^2/3+...]
\end{equation}

\noindent when $n=1/2$.  The problem is that as we integrate backwards ($t<t_0$) the last term in
equations (\ref{N2}) or (\ref{N3}) grows rapidly, and the corresponding vacuum energy can become
dynamically dominant at critical cosmological epochs, particularly nucleosynthesis, recombination
and galaxy formation.  This must not be allowed to happen, and the singular component must be
eliminated (Ratra and Peebles 1988).  The problem is not peculiar to the particular potential
(\ref{V}), but is encountered whenever we use equation (\ref{K}) with an explicit potential
as a model of vacuum energy; in earlier work the problem was avoided by setting initial
conditions at $t=0$: $\dot\phi(0)=0$ and $\phi(0)$ fine-tuned to give the desired current 
conditions, as in Jackson (1998), Jackson and Dodgson (1998) and Weller and Albrecht (2002).
There is a possible degeneracy here, in that strictly speaking there is an allowable range of non-zero
values of $|\dot\phi(0)|<<\omega_{\hbox{\scriptsize{C}}}|\phi(0)|$; however, the constraints imposed
by the above-mentioned astrophysical are constraints are severe, and within the allowable range the
the results to be presented here are not noticeably dependent on the value of $|\dot\phi(0)|$.    
In the present context we find the above-mentioned approach somewhat cumbersome, and have adopted
an alternative means of achieving the same ends, as follows.

We are almost certain that the expansion is accelerating, which in this context means 
that $\phi$ is close to its slow-roll value and is falling slowly towards the minimum
in $V(\phi)$ at $\phi=0$; during this phase we expect $\dot\phi$ to be given by

\begin{equation}\label{O}
\dot\phi=-{\omega_{\hbox{\scriptsize{C}}}^2\phi \over 3H}
\end{equation}

\noindent which as an initial condition on $v$ would be

\begin{equation}\label{P}
v(t_0)={-\omega_{\hbox{\scriptsize{C}}}^2u(t_0)/3}.
\end{equation}

\noindent Equations (\ref{M}) and (\ref{P}) together give

\begin{equation}\label{Q}
u(t_0)=\left({\Omega_\phi \over \omega_{\hbox{\scriptsize{C}}}^2+\omega_{\hbox{\scriptsize{C}}}^4/9}\right)^{1/2}.
\end{equation}

We have tried expressions (\ref{P}) and (\ref{Q}) as starting values, and find them to
be wanting.  Instead we must retain a degree of freedom, and use

\begin{equation}\label{R}
v(t_0)={-\omega_{\hbox{\scriptsize{C}}}^2u(t_0)/\gamma}
\end{equation}

\noindent and

\begin{equation}\label{S}
u(t_0)=\left({\Omega_\phi \over \omega_{\hbox{\scriptsize{C}}}^2+\omega_{\hbox{\scriptsize{C}}}^4/\gamma^2}\right)^{1/2}.
\end{equation}

\noindent The procedure is to integrate back to a suitably high fixed redshift $z_{\hbox{\scriptsize{f}}}$
(depending on the application, for example $z_{\hbox{\scriptsize{f}}}=z_{\hbox{\scriptsize{r}}}=1088$
to determine $d_{\hbox{\scriptsize{A}}}(z_{\hbox{\scriptsize{r}}})$), and to find $\Omega_{\hbox{\scriptsize{m}}}
(z_{\hbox{\scriptsize{f}}})$ as a function of the variable $\gamma$.  The correct value of $\gamma$ is the one which maximizes $\Omega_{\hbox{\scriptsize{m}}}(\gamma,z_{\hbox{\scriptsize{f}}})$.  As an indication of the necessity
of this undertaking, we quote figures for a model with current values $\Omega_{\hbox{\scriptsize{m}}}=0.3$, $\Omega_\phi=0.7$; with $\omega_{\hbox{\scriptsize{C}}}=0$ we find $\Omega_{\hbox{\scriptsize{m}}}(\gamma,10)=0.9956$, the balance comprising scalar energy and radiation; with $\omega_{\hbox{\scriptsize{C}}}=1$ we find
$\Omega_{\hbox{\scriptsize{m}}}(3,10)=0.1140$, whereas the optimization procedure gives
$\Omega_{\hbox{\scriptsize{m}}}(3.59,10)=0.9952$.  In other words the slow-roll value $\gamma=3$ gives
a universe in which the scalar component becomes increasingly dominant as we move to earlier epochs,
whereas for the correct value $\gamma=3.59$ the opposite is the case.

\begin{figure}[here]
\begin{center}
\includegraphics[width=12.5cm] {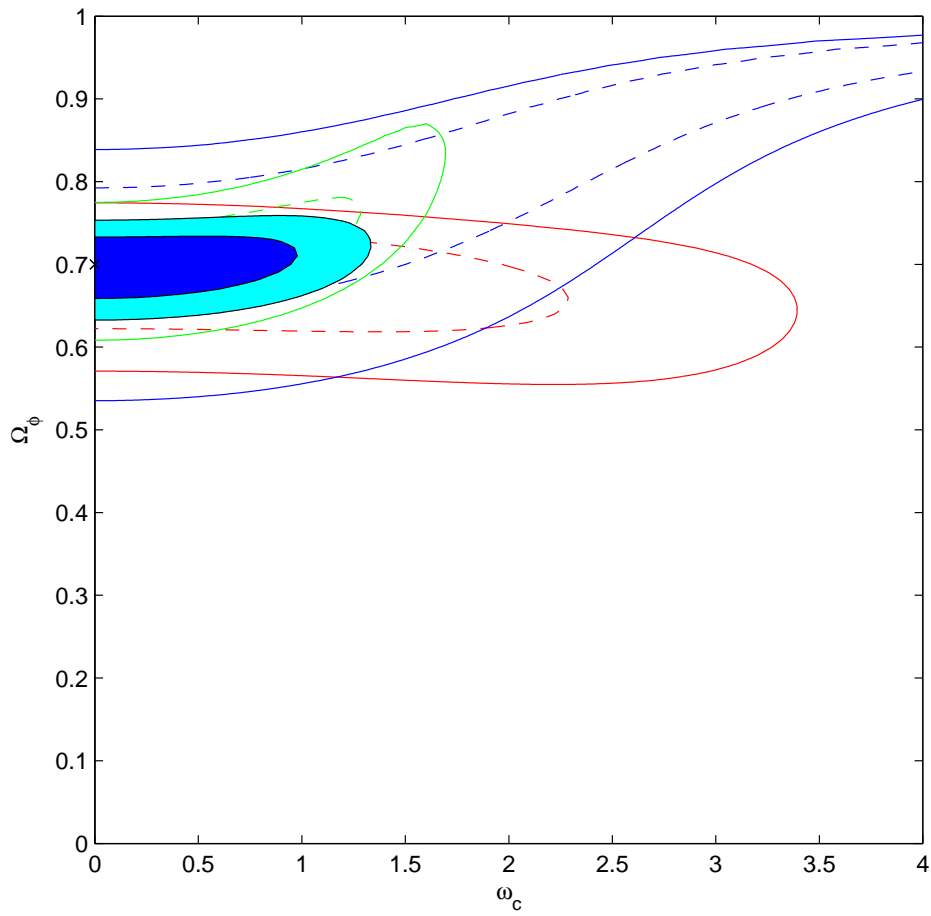}
\end{center}
\caption{\label{FigwC_sigmas} Confidence regions in the $\omega_{\hbox{\scriptsize{C}}}$--$\Omega_\phi$ plane,
95\% and 68\%, where $\omega_{\hbox{\scriptsize{C}}}$ is Compton frequency in Hubble units;
as determined by the three data sets: red=milliarcsecond radio-sources, blue=acoustic horizon,
green=Type Ia supernovae; the filled areas are joint confidence regions for all three.
Flat universes with evolving scalar vacuum energy governed by a potential
$V(\phi)=\omega_{\hbox{\scriptsize{C}}}^2\phi^2/2$.}
\end{figure}

Figure \ref{FigwC_sigmas} shows 68\% and 95\% confidence regions in the $\omega_{\hbox{\scriptsize{C}}}$--$\Omega_\phi$ plane, to be discussed below.

\section{Conclusions}

In isolation, none of the techniques currently used to constrain cosmological parameters
give precise numerical answers; at best we get answers to profound qualitative questions,
relating particularly to acceleration and deceleration.  It is only in combination
that such techniques give meaningful numerical results.  Nevertheless, we believe that
our figures relating to milliarcsecond radio-sources are the most accurate single-technique
ones to date, fully justifying the expectations expressed in the last paragraph of the
Introduction; those relating to the joint samples MAS+Ach, SN+AcH and MAS+SN+Ach
(Tables \ref{Table295} and \ref{Table268}) are close to the best currently available
(see for example Tonry \etal 2003; Riess \etal 2004; Astier \etal 2006).
Figures \ref{FigwC_sigmas} shows that there is no evidence here for dynamical vacuum energy,
in that the data prefer the value $\omega_{\hbox{\scriptsize{C}}}=0$.
However, the data are compatible with a current rate-of-change corresponding to
$\omega_{\hbox{\scriptsize{C}}}=1.34$ times the Hubble rate, which is thus faster than
the latter and must be regarded as rapid.  If the vacuum energy really is constant,
the best we will ever do is to place limits on its variation, which is a potential dilemma for physics,
because the smallest variation would suggest something like a scalar model, whereas true constancy
would imply something more fundamental.

\begin{figure}[here]
\begin{center}
\includegraphics[width=12.5cm] {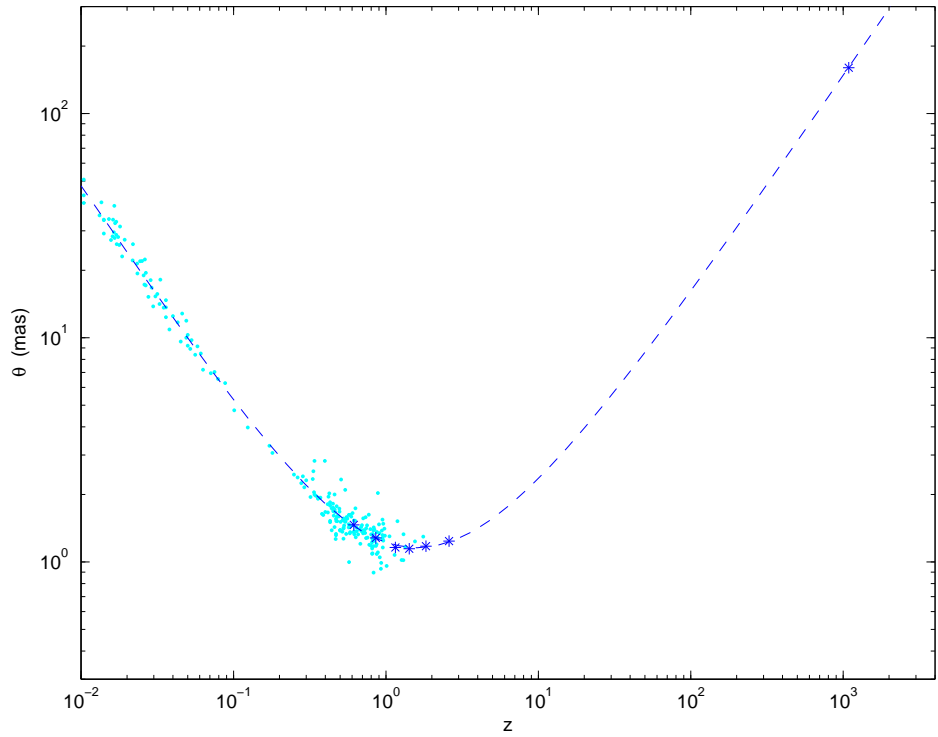}
\end{center}
\caption{\label{Fig_theta_z_sigmas} Angular-diameter/redshift diagram; each lower blue star is a bin
containing c. 77 milliarcsecond radio-sources (see text); the error bars are too small to be seen
on this scale, amounting to $\pm 0.006$ in $\log\theta$. Each cyan point corresponds to a
supernova transformed into a mock milliarcsecond source.  The top blue star represents the 
acoustic horizon; the error bar is $\pm 0.0120$ in $\log\theta$, corresponding to $h=0.72\pm 0.05$.}
\end{figure}

\begin{figure}[here]
\begin{center}
\includegraphics[width=12.5cm] {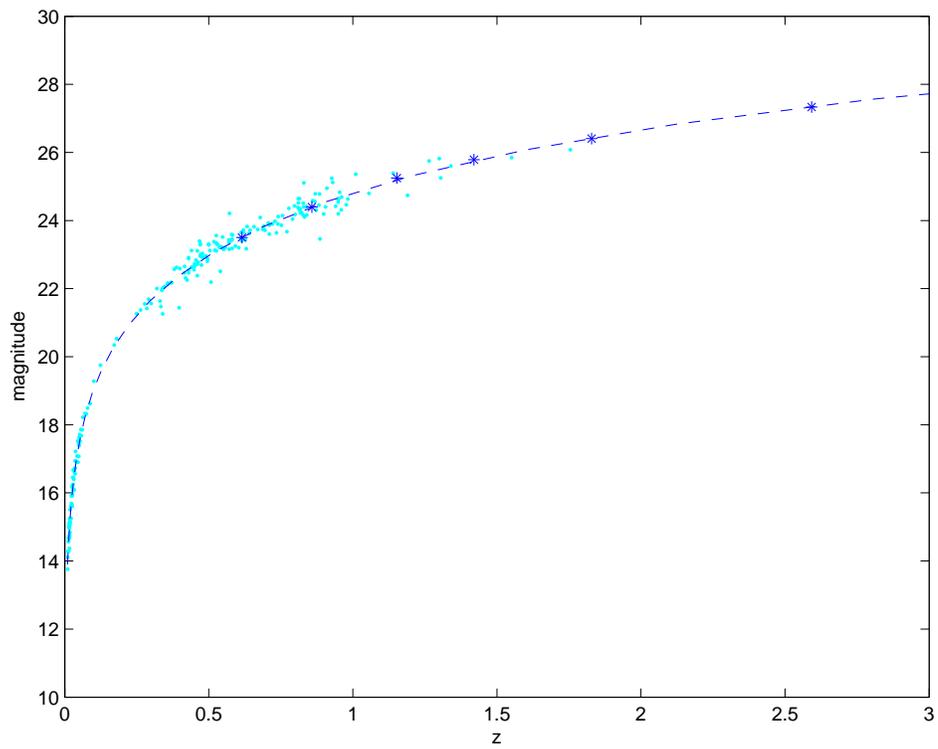}
\end{center}
\caption{\label{Fig_m_z_sigmas} Magnitude/redshift diagram; each cyan point is a supernova.
Each blue star represents a bin containing c. 77 milliarcsecond radio-sources (see text);
the error bars are too small to be seen on this scale, amounting to $\pm 0.03$ magnitudes.
}
\end{figure}

Figures \ref{Fig_theta_z_sigmas} and \ref{Fig_m_z_sigmas} are graphical summaries of the data used
in this investigation.  The dashed blue curve is for a model with $\Omega_{\hbox{\scriptsize{m}}}=0.304$,
$\Omega_\Lambda=0.693$.  Mock radio sources were generated from supernovae using
the angular-diameter distances defined by each apparent magnitude and a fixed
absolute magnitude $M=-19.28+5\log_{10}(h/0.7)$, and a fixed linear size $d=6.839h^{-1}$ pc;
$M$ and $d$ are the best-fitting values for the samples SN and MAS respectively,
when $\Omega_{\hbox{\scriptsize{m}}}$ and $\Omega_\Lambda$ are fixed at the above values.
Mock supernovae were generated from radio sources in similar fashion.
The mock radio source corresponding to the acoustic horizon was generated using $d$,
and an angular diameter distance defined by its observed angular size and the value of
of $l_{\hbox{\scriptsize{AcH}}}$ given by equations (\ref{A}) to (\ref{D}), with $h=0.72$,
$\Omega_{\hbox{\scriptsize{b}}}h^2=0.024$ and $\Omega_{\hbox{\scriptsize{m}}}=0.304$.
With these definitions the data points reflect the observational spread.

{\it Postscript}.  As this paper was about to be submitted the WMAP three year results became available
(Spergel \etal 2006; Hinshaw \etal 2006).  We have had a quick look at the new data, to assess their
impact on the figures presented above.  According to the scheme adopted above we should use the new
best-fitting flat $\Lambda$CDM values $h=0.73\pm 0.03$, $\Omega_{\hbox{\scriptsize{b}}}h^2=0.0223$
and $l_{\hbox{\scriptsize{peak}}}=220.7$.  The main effect of these changes is to move the AcH region
in Figure \ref{FigOmega_sigmas} slightly to the left, so that the intersection with the flat line, previously
at $\Omega_{\hbox{\scriptsize{m}}}=0.271$, is now at $\Omega_{\hbox{\scriptsize{m}}}=0.240$.
This is in line with the WMAP three year flat $\Lambda$CDM values, alone and in combination with a
number of other astronomical data sets (Spergel \etal 2006); the exceptions are WMAP+supernovae
(the gold subset of Riess \etal (2004)) and WMAP+weak lensing data, which continue to favour somewhat higher
values.  Our new figures are summarized in Table \ref{Table568}; with the exception of the curvature
parameter $K$, these are virtually the same as those in Table \ref{Table268}; the new values of $K$
indicate a marginally open Universe; the corresponding error bars are smaller, because the blue region
in Figure \ref{FigOmega_sigmas} is now narrower, but the conclusion $K<0$ is not yet statistically significant.

\begin{table}[here]
\caption{\label{Table568} Best-fitting cosmological parameters, 68\% confidence limits,
taking into account the WMAP three year results; MAS=milliarcsecond radio-sources,
AcH=acoustic horizon, SN=Type Ia supernovae.}
\begin{indented}
\item[]\begin{tabular}{@{}cccc}
\br
Sample & $\Omega_{\hbox{\scriptsize{m}}}$ & $\Omega_\Lambda$ & K\\
\mr
MAS+AcH &    $0.311 +0.037/\!-0.034 $&$ 0.673 +0.046/\!-0.050 $&$ -0.016 +0.017/\!-0.017$ \\
SN+AcH &     $0.297 +0.033/\!-0.031 $&$ 0.690 +0.039/\!-0.042 $&$ -0.013 +0.012/\!-0.013$ \\  
MAS+SN+AcH & $0.303 +0.025/\!-0.022 $&$ 0.684 +0.030/\!-0.033 $&$ -0.013 +0.012/\!-0.012$ \\ 
\br
\end{tabular}
\end{indented}
\end{table}

\bigbreak
\noindent{\bf Acknowledgements}
\medbreak

This research has made use of the NASA/IPAC Extragalactic Database (NED) which is operated
by the Jet Propulsion Laboratory, California Institute of Technology, under contract with the
National Aeronautics and Space Administration.

\newpage
\References

\item[] Aldering G, 2005 {\it New Astron. Reviews} {\bf 49} 346
\item[] Astier \etal, 2006 {\it Astron. Astrophys.} {\bf 447} 31  
\item[] Altavilla G \etal, 2004 {\it Mon. Not. R. Astron. Soc.} {\bf 349} 1344
\item[] Balbi A \etal, 2000 { \it Astrophys. J.} {\bf 545} L1
\item[] Barthel P D and Miley G K, 1988 {\it  Nature} {\bf 333} 319
\item[] Bassett B A, Corasaniti P S and Kunz M, 2004 { \it Astrophys. J.} {\bf 617} L1
\item[] Buchalter A, Helfand D J, Becker R H and White R L, 1998 {\it Astrophys. J.}
        {\bf 494} 503
\item[] Chiba T and Nakamura T, 2000 {\it Phys. Rev.} {\bf D62} 121301 
\item[] Condon J J, Cotton W D, Greisen E W, Yin Q F, Perley R A, Taylor G B and Broderick J J,
        1998 {\it Astronom. J.} {\bf 115} 1693
        (http://www.cv.nrao.edu/nvss/) 
        (http://www.mrao.cam.ac.uk/surveys/nrao/NVSS/NVSS.html)
\item{} Cooray A R and Huterer D, 1999 {\it Astrophys. J.} {\bf 513} L95          
\item{} Daly R A, 1994 {\it Astrophys. J.} {\bf 426} 38 
\item[] Daly R A and Djorgovski S G, 2003 {\it Astrophys. J.} {\bf 597} 9
\item[] Daly R A and Djorgovski S G, 2004 {\it Astrophys. J.} {\bf 612} 652
\item[] de Bernardis P \etal, 2000 {\it  Nature} {\bf 404} 955
\item[] D'Oderico S, Dessuages-Zavadsky M and Molaro P, 2001 {\it Astron. Astrophys.} {\bf 368} L21
\item[] Fanaroff B L and Riley J M, 1974 {\it Mon. Not. R. Astron. Soc.} {\bf 167} 31P  
\item[] Freedman W L \etal, 2001 {\it Astrophys. J.} {\bf 553} 47
\item[] Gerke B F and Efstathiou G, 2002 {\it Mon. Not. R. Astron. Soc.} {\bf 335} 33
\item[] Gurvits L I, 1994 {\it Astrophys. J.} {\bf 425} 442
\item[] Guy J, Astier P, Nobili S, Regnault N and Pain R, 2005 {\it Astron. Astrophys.} {\bf 443} 781
\item[] Hanany S \etal, 2000 {\it Astrophys. J.} {\bf 545} L5
\item[] Hinshaw G \etal, 2003 {\it Astrophys. J. Suppl.} {\bf 148} 135
\item[] Hinshaw G \etal, 2006 Preprint astro-ph/0603451
\item[] Huterer D and Turner M S, 1999 {\it Phys. Rev.} {\bf D60} 081301
\item[] Huterer D and Turner M S, 2001 {\it Phys. Rev.} {\bf D64} 123527
\item[] Huterer D and Starkman G, 2003 {\it Phys. Rev. Lett.} {\bf 90} 031301  
\item[] Hu W and Sugiyama N, 1995 {\it Phys. Rev.} {\bf D51} 2599 
\item[] Jackson J C, 1973 {\it Mon. Not. R. Astron. Soc.} {\bf 162} 11P
\item[] Jackson J C and Dodgson M, 1997 {\it Mon. Not. R. Astron. Soc.} {\bf 285} 806
\item[] Jackson J C, 1998 {\it Mon. Not. R. Astron. Soc.} {\bf 296} 619
\item[] Jackson J C and Dodgson M, 1998 {\it Mon. Not. R. Astron. Soc.} {\bf 297} 923
\item[] Jackson J C, 2004 {\it JCAP} {\bf 11} 7
\item[] Jassal H K, Bagla J S and Padmanabhan T, 2006 Preprint astro-ph/0601389
\item[] J\H onnson J, Goobar A, Amanullah R and Bergstr\H om L, 2004 {\it JCAP} {\bf 09} 7
\item[] Kellermann K I, 1993 {\it Nature} {\bf 361} 134  
\item[] Kirkman D, Tytler D, Suzuki N, O'Meara J and Lubin D, 2003 {\it Astrophys. J. Suppl.} {\bf 149} 1
\item[] K\H uhr H, Witzel A, Pauliny-Toth I I K and Nauber U, 1981 {\it Astron. Astrophys.} {\bf 45} 367
\item[] Legg T H, 1970 {\it  Nature} {\bf 226} 65
\item[] Linder E V, 2003  {\it Phys. Rev. Lett.} {\bf 90} 91031
\item[] Linder E V, 2004  {\it Phys. Rev.} {\bf D70} 061302
\item[] Masson C R, 1980 {\it Astrophys. J.} {\bf 242} 8
\item[] Maor I, Brustein R and Steinhardt P J, 2001 {\it Phys. Rev. Lett.} {\bf 86} 6
\item[] Maor I, Brustein R, McMahon J and Steinhardt P J, 2002 {\it Phys. Rev.} {\bf D65} 123003
\item[] Miley G K, 1971 {\it Mon. Not. R. Astron. Soc.} {\it 152} 477
\item[] Mosoni L, Frey S, Gurvits L I, Garrett M A, Garrington S T and Tsvetanov Z I, 2006
        {\it Astron. Astrophys.} {\bf 445} 413
\item[] Mukhanov V, 2004 {\it Int. J. Theor. Phys.} {\bf 43} 623
\item[] Nilsson K, Valtonen  J, Kotilainen J and Jaakkola T, 1993
        {\it Astrophys. J.} {\bf 413} 453
\item[] Parkes Catalogue, 1990, Australia Telescope National Facility, Wright A and Otrupcek R (Eds)
        (http://www.parkes.atnf.csiro.au/research/surveys/pkscat90.html)
\item[] Perlmutter S. \etal, 1997 {\it Astrophys. J.} {\bf 483} 565
\item[] Perlmutter S. \etal, 1999 {\it Astrophys. J.} {\bf 517} 565
\item[] Pettini M and Bowen D V, 2001 {\it Astrophys. J.} {\bf 560} 41
\item[] Preston R A, Morabito D D, Williams J G, Faulkner J, Jauncey D L,
        and Nicolson G, 1985 {\it Astronom. J.} {\bf 90} 1599 (P85)
\item[] Ratra B and Peebles P J E, 1988 {\it Phys. Rev.} {\bf D37} 3406
\item[] Richter G M, 1973 {\it Astrophys. Lett.} {\bf 13} 63
\item[] Riess A G \etal, 1998 {\it Astronom. J.} {\bf 116} 1009
\item[] Riess A G \etal, 2004 {\it Astrophys. J.} {\bf 607} 665 
\item[] Riess A G, Li W, Stetson P B, Filippenko A V, Jha S, Kirshner R P, Challis P M, Garnavich P M and Chornock R,
        2005 {\it Astrophys. J.} {\bf 627} 579
\item[] Schmidt B P \etal, 1998 {\it Astrophys. J.} {\bf 507} 46
\item[] Seljak U and Zaldarriaga M, 1996 {\it Astrophys. J.} {\bf 469} 437
\item[] Singal A K, 1988 {\it Mon. Not. R. Astron. Soc.} {\bf 233} 87
\item[] Spergel D N \etal, 2003 {\it Astrophys. J. Suppl.} {\bf 148} 175
\item[] Spergel D N \etal, 2006 Preprint astro-ph/0603449
\item[] Thompson A R, Moran J M and Swenson G W Jr, 1986
        {\it Interferometry and Synthesis in Radio Astronomy} (New York: Wiley) p 13
\item[] Tonry J L \etal, 2003 {\it Astrophys. J.} {\bf 594} 1   
\item[] V\'eron-Cetty M-P and V\'eron P, 1991 {\it A Catalogue of Quasars and Active Glactic
        Nuclei} (5th Edition ESO Science Report No. 10)
\item[] Wang Y and Freese K, 2006 {\it Phys. Lett.} {\bf B632} 449
\item[] Weller J and Albrecht A, 2002 {\it Phys. Rev.} {\bf D65} 103512
\item[] Wilkinson P N, Polatidis A G, Readhead A C S, Xu W and Pearson T J, 1994
        {\it Astrophys. J.} {\bf 432} L87

\endrefs

\newpage
\section{Figures}

\newpage

\end{document}